\documentclass[runningheads]{llncs}
\usepackage{graphicx}
\usepackage{url}
\usepackage{hyperref}
\usepackage{bytefield}
\usepackage{xspace}
\usepackage{enumitem}
\usepackage{multirow}

\graphicspath{{figures/}}

\newcommand{\icmp}{ICMP\xspace}
\newcommand{\vfour}{IPv4\xspace}
\newcommand{\eg}{e.g.,\ }
\newcommand{\ie}{i.e.,\ }
\newcommand{\etal}{et al.\ }
\providecommand{\e}[1]{\ensuremath{\times 10^{#1}}}

\newcommand{\punkt}[1]{\item\textbf{#1}:}
\newcommand{\orig}{\texttt{orig\_ts}}
\newcommand{\recv}{\texttt{recv\_ts}}
\newcommand{\xmit}{\texttt{xmit\_ts}}
\newcommand{\reqXmit}{\texttt{request\textsubscript{xmit\_ts}}}
\newcommand{\reqRecv}{\texttt{request\textsubscript{recv\_ts}}}
\newcommand{\sundial}{\texttt{sundial}\xspace}
\newcommand{\tz}{timezone\xspace}
\newcommand{\tzp}{timezones\xspace}
\newcommand{\Tz}{Timezone\xspace}

\newcounter{magicrownumbers}
\newcommand\rownumber{\stepcounter{magicrownumbers}\arabic{magicrownumbers}}

\begin{document}

\title{Sundials in the Shade}
\subtitle{An Internet-wide Perspective on ICMP Timestamps}

\author{Erik C. Rye 
\and Robert Beverly
}
\institute{Naval Postgraduate School, Monterey, CA \\
\email rye@cmand.org, rbeverly@nps.edu}

\authorrunning{E. C. Rye and R. Beverly}
\maketitle
\begin{abstract}
ICMP timestamp request and response packets have been standardized for
nearly 40 years, but have no modern practical application, having been
superseded by NTP.  However, ICMP timestamps are not deprecated,
suggesting that while hosts must support them, little attention is
paid to their implementation and use.  In this work, we perform active
measurements and find 2.2 million hosts on the Internet responding to ICMP
timestamp requests from over 42,500 unique autonomous systems.  We develop a
methodology to classify timestamp responses, and find 13 distinct
classes of behavior.  Not only do these behaviors enable a new
fingerprinting vector, some behaviors leak important information about the host
\eg OS, kernel version, and local \tz.

\end{abstract}

\section{Introduction}
\label{sec:intro}

The Internet Control Message Protocol (\icmp) is part of the original
Internet Protocol specification (\icmp is IP protocol number one), and has
remained largely unchanged since RFC 792~\cite{rfc792}.  Its primary function is to
communicate error and diagnostic information; well-known uses today
include \icmp echo to test for reachability (\ie \texttt{ping}), \icmp
time exceeded to report packet loops (\ie \texttt{traceroute}), and
\icmp port unreachable to communicate helpful information to the
initiator of a transport-layer connection.  Today, 27 \icmp types are
defined by the IESG, 13 of which are deprecated~\cite{iana-icmp}.

Among the non-deprecated \icmp messages are timestamp (type 13) and
timestamp reply (type 14).  These messages, originally envisioned to
support time synchronization and provide one-way delay
measurements~\cite{rfc778}, contain three 32-bit time values that
represent milliseconds (ms) since midnight UTC.  Modern clock
synchronization is now performed using the Network Time
Protocol~\cite{rfc5905} and \icmp timestamps are generally regarded as
a potential security vulnerability~\cite{CVE-1999-0524} as they can
leak information about a remote host's clock.  
Indeed, Kohno \etal demonstrated in 2005 the potential to identify
individual hosts by variations in their clock
skew~\cite{kohno2005remote}, while \cite{desmond2008identifying} and
\cite{cristea2013fingerprinting} show similar discriminating power
when fingerprinting wireless devices.  

In this work, we reassess the extent to which Internet hosts respond
to \icmp timestamps.  Despite no legitimate use for \icmp
timestamps today, and best security practices that recommend blocking
or disabling these timestamps, we receive timestamp responses from 
2.2
million \vfour hosts in 42,656
distinct autonomous systems (approximately 15\% of the hosts queried) 
during 
a large-scale
measurement campaign in September and October 2018.
In addition to characterizing this unexpectedly
large pool of responses, we seek to better understand \emph{how} hosts respond.
Rather than focusing on clock-skew fingerprinting, we instead make the following
primary contributions: 
\begin{enumerate}  
  \item The first Internet-wide survey of \icmp timestamp
  support and responsiveness.
  \item A taxonomy of \icmp timestamp response behavior, and a 
  methodology to classify responses.
  \item Novel uses of \icmp timestamp responses, including
  fine-grained operating system fingerprinting and coarse
  geolocation.
\end{enumerate}  

\section{Background and Related Work}

\begin{figure}[t]
  \centering
\begin{bytefield}[bitwidth=1em,bitheight=1.2em]{32}
  \bitheader{0-31} \\
  \bitbox{8}{\texttt{type=13/14}} &\bitbox{8}{\texttt{code=0}}
  & \bitbox{16}{\texttt{checksum}} \\
  \bitbox{16}{\texttt{id}}
  & \bitbox{16}{\texttt{sequence}} \\
  \bitbox{32}{\orig}\\
  \bitbox{32}{\recv}\\
  \bitbox{32}{\xmit}
\end{bytefield}
 \vspace{-4mm}
 \caption{\icmp Timestamp Message Fields}
 \label{fig:ts1}
 \vspace{-6mm}
\end{figure}

Several TCP/IP protocols utilize timestamps, and significant prior
work has examined TCP timestamps in the context of
fingerprinting~\cite{kohno2005remote}.  TCP timestamps have since been
used to infer whether IPv4 and IPv6 server addresses map to the same
physical machine in~\cite{beverly2015server} and combined with clock
skew to identify server ``siblings'' on a large scale
in~\cite{scheitle2017large}.

In contrast, this work focuses on \icmp timestamps.
Although originally intended to support time
synchronization~\cite{rfc778}, \icmp timestamps have no modern
legitimate application use (having been superseded by NTP).  Despite
this, timestamps are not deprecated~\cite{iana-icmp}, suggesting that
while hosts must support them, little attention is paid to their
implementation and use.  

Figure~\ref{fig:ts1} depicts the structure of timestamp request (type
13) and response (type 14) \icmp messages. The 16-bit identifier and
sequence values enable responses to be associated with requests.  
Three four-byte fields are defined: the
\emph{originate timestamp} (\orig), \emph{receive timestamp} (\recv),
and \emph{transmit timestamp} (\xmit).  Per RFC792~\cite{rfc792}, timestamp
fields encode milliseconds (ms) since UTC midnight unless the most
significant bit is set, in which case the field may be a
``non-standard''
value.  The originator of timestamp
requests should set the originate timestamp using her own clock; the value of the receive
and transmit fields for timestamp requests is not specified in the RFC.

To respond to an \icmp timestamp request, a host simply copies the request
packet, changes the \icmp type, and sets the receive and transmit time
fields.  The receive time indicates when the request
was received, while the transmit time indicates when the reply was
sent.

Several prior research works have explored
\icmp timestamps, primarily for fault diagnosis and fingerprinting.
Anagnostakis \etal found in 2003 that 93\% of the approximately 400k
routers they probed responded to \icmp timestamp requests, and
developed a tomography technique using \icmp timestamps to measure
per-link one-way network-internal delays~\cite{anagnostakis2003cing}.
Mahajan \etal leveraged and expanded the use of \icmp timestamps to
enable user-level Internet fault and path diagnosis
in~\cite{mahajan2003user}.

Buchholz and Tjaden leveraged \icmp timestamps in the context of
forensic reconstruction and correlation~\cite{buchholz2007brief}.
Similar to our results, they find a wide variety of clock behaviors.
However, while they probe $\sim$8,000 web servers, we perform
an Internet-wide survey including 2.2M hosts more than a decade later,
and demonstrate novel fingerprinting and geolocation uses of \icmp
timestamps. 

Finally, the \texttt{nmap} security scanner~\cite{nmap} uses \icmp timestamp
requests, in addition to other protocols, during host discovery for
non-local networks 
in order to circumvent firewalls and blocking. \texttt{nmap} sets
the request originate timestamp to zero by default,
in violation of the standard~\cite{rfc792}
(though the user can manually specify a timestamp).  Thus, \icmp timestamp
requests with zero-valued origination times provide a signature of
\texttt{nmap} scanners searching for live hosts.
While 
\texttt{nmap} uses \icmp timestamps for liveness testing, it does not use
them for operating system detection as we do in this work.

To better understand the prevalence of \icmp timestamp scanners, we
analyze 240 days of traffic arriving at a /17 network telescope.
We observe a total of 413,352 timestamp messages, 93\% of which are
timestamp requests.  Only 33 requests contain a non-zero
originate timestamp, suggesting that the remainder (nearly 100\%) are
\texttt{nmap} scanners.  The top 10 sources account for more than 86\% of 
the requests we observe, indicating a relatively small
number of active Internet-wide scanners.

\section{Behavioral Taxonomy}
\label{sec:taxonomy}

During initial probing, we found significant variety
in timestamp responses.  Not only do structural differences exist in the
implementation of \cite{rfc792} by timestamp-responsive routers and
end systems (\eg little- vs big-endian), they
also occur relative to how the device counts time (\eg
milliseconds vs.\ seconds), the device's reference point (\eg
UTC or local time), whether the reply is a function of
request parameters, and even whether the device is keeping time at all.

\begin{table}[t]
 \centering
 \caption{\icmp Timestamp Classification Fingerprints}
 \label{tab:taxonomy}
 \vspace{-3mm}
 \scriptsize{
 \begin{tabular}{|l|l|c|c|c|c|c|}\hline
   &       & \textbf{Request} & \multicolumn{3}{|c|}{\textbf{Response}} \\\hline
   \textbf{Num} & \textbf{Class} & \texttt{cksum} & \orig        & \recv   & \xmit \\\hline\hline
   \rownumber & Normal & valid &- & $\ne \xmit, \ne 0$ & $\ne0$ \\\hline
   \rownumber & Lazy & valid &- & = \xmit    & $\ne0$ \\\hline
   \rownumber & Checksum-Lazy & bad & - & - & - \\\hline
   \rownumber & Stuck &valid &- & const & const \\\hline
   \rownumber & Constant 0 &valid & - & 0  & 0 \\\hline  
   \rownumber & Constant 1 &valid & - & 1  & 1 \\\hline
   \rownumber & Constant LE 1 &valid &- & $ htonl(1)$  & $htonl(1)$ \\\hline
   \rownumber & Reflection &valid & - & \reqRecv & \reqXmit \\\hline
   \rownumber & Non-UTC &valid & - & $>2^{31}-1$ & $>2^{31}-1$\\\hline
   \rownumber & \Tz &valid & - & $|\recv-\orig|\%\left(3.6\e{6}\right) < 200\ ms$ & -\\\hline
   \rownumber & Little Endian &valid & - & $|htonl(\recv) - \orig| < 200\ ms$ & - \\\hline
   \rownumber & Linux $htons()$ Bug &valid & - & $ \% 2^{16} = 0$ & $\% 2^{16} = 0$ \\\hline
   \rownumber & Unknown  &valid  & - & - & - \\\hline
 \end{tabular}
 }
 \vspace{-6mm}
\end{table}

\subsection{Timestamp Implementation Taxonomy}
\label{sec:taxonomy:structural}

Table~\ref{tab:taxonomy} provides an exhaustive taxonomy of the
behaviors we observe; we term these the \icmp timestamp \emph{classifications}. Note that
this taxonomy concerns
only the \emph{implementation} of the timestamp response, rather than
whether the responding host's timestamp values are correct. 

\begin{itemize}[wide,labelwidth=!,labelindent=0pt]
  \punkt{Normal} 
    Conformant to \cite{rfc792}.
    Assuming more than one ms of
    processing time, the receive and transmit timestamps should be not equal,
    and both should be nonzero except at midnight UTC. 
  \punkt{Lazy} 
    Performs a single time lookup and sets both receive and transmit
    timestamp fields to the same value.
    A review of current Linux and FreeBSD kernel source code reveals
    this common lazy implementation~\cite{freebsdicmp,linuxicmp}.
  \punkt{Checksum-Lazy} 
    Responds to timestamp
    requests even when the \icmp checksum is incorrect.
  \punkt{Stuck} 
    Returns
    the same value in the receive and transmit timestamp fields regardless of
    the input sent to it and time elapsed between probes. 
  \punkt{Constant 0, 1, Little-Endian 1} 
    A strict
    subset of ``stuck'' that always returns a small
    constant value in the receive and transmit timestamp fields.
  \punkt{Reflection} 
    Copies the
    receive and transmit timestamp fields from the timestamp request into the
    corresponding fields of the reply message\footnote{We find
    no copying of originate timestamp into the reply's
    receive or transmit fields.}. 
  \punkt{Non-UTC} 
    Receive and
    transmit timestamp values with the most significant bit set.
    As indicated in \cite{rfc792}, network devices that are
    unable to provide a timestamp with respect to UTC midnight or in
    ms may use an alternate time source, provided that the high order
    bit is set. 
  \punkt{Linux \emph{htons()} Bug} 
    Certain versions of the Linux kernel (and Android) contain a flawed
    \icmp timestamp implementation where replies are truncated to a
    16-bit value; see Appendix A for details.
  \punkt{Unknown} Any reply not
    otherwise classified.
\end{itemize}

\subsection{Timekeeping Behavior Taxonomy}
\label{sec:timekeeping}
We next categorize the types of timestamp responses we
observe by what the host is measuring and what they are measuring in relation
to. 
\begin{itemize}[wide,labelwidth=!,labelindent=0pt]
  \punkt{Precision} Timestamp reply fields should encode ms to be
   conformant, however some implementations encode seconds.
  \punkt{UTC reference} Conformant to the RFC; receive and transmit 
     timestamps encode ms since midnight UTC.
  \punkt{\Tz} 
    Replies with receive and transmit timestamps in ms relative to
    midnight in the device's local \tz, rather than UTC midnight. 
  \punkt{Epoch reference} Returned timestamps encode time in seconds
     relative to the Unix epoch time.
  \punkt{Little-Endian} 
    Receive and transmit timestamps containing a correct timestamp when
    viewed as little-endian four-byte integers. 
\end{itemize}

\section{Methodology}
\label{sec:methodology}

We develop \sundial, a packet prober that implements the methodology
described herein to elicit timestamp responses that permit behavioral
classification.  \sundial is written in C 
and sends raw IP packets in
order to set specific IP and ICMP header fields, while targets
are randomized to distribute load. We have since ported \sundial to
a publicly available ZMap~\cite{durumeric2013zmap} module~\cite{sundial}.

Our measurement survey consists of probing 14.5 million \vfour
addresses\footnote{As IPv6 does not support timestamps in ICMPv6, we
study IPv4 exclusively.} of the August 7, 2018 ISI hitlist, which
includes one address per routable /24 network~\cite{fan2010selecting}.
We utilize two vantage points connected to large academic university 
networks named after their respective locations:
``Boston'' and ``San Diego.''  
Using \sundial, we
elicit \icmp timestamp replies from $\sim$2.2 million unique IPs.

This section first describes \sundial's messages and methodology, 
then our ground truth validation.  We then discuss
ethical concerns and precautions undertaken in this study.

\subsection{\sundial Messages}
\label{sec:requesttypes}

In order to generate and categorize each of the response behaviors,
\sundial transmits four distinct types of \icmp timestamp requests.
Both of our vantage points have their time NTP-synchronized
to stratum 2 or better servers.  Thus time is ``correct'' on our
prober relative to NTP error.
\begin{enumerate}[wide,labelwidth=!,labelindent=0pt]
  \punkt{Standard} 
    We fill the
    originate timestamp field with the correct ms from UTC
    midnight, zero the receive and transmit timestamp fields, and place the
    lower 32 bits of the MD5 hash of the destination IP address and originate
    timestamp into the identifier and sequence number fields. The 
    hash
    permits detection of
    destinations or
    middleboxes that tamper with the originate timestamp, identifier, or
    sequence number.
  \punkt{Bad Clock} 
    We zero the receive and transmit fields of the request,
    choose an identifier and sequence number, and compute the MD5 hash of the
    destination IP address together with the identifier and sequence number. The
    lower 32 bits of the hash are placed in the originate timestamp.
    This hash again
    provides the capability to detect modification of the
    reply. 
  \punkt{Bad Checksum} 
    The correct time in ms since
    UTC midnight is placed in the originate field, the receive and transmit
    timestamps are set to zero, and the identifier and sequence number fields
    contain an encoding of the destination IP address along with the originate
    timestamp.
    We deliberately choose a random, incorrect
    checksum and place it into the \icmp timestamp request's checksum field.
    This timestamp message should appear corrupted to the destination, and a
    correct \icmp implementation should discard it. 
  \punkt{Duplicate Timestamp} 
    The receive and transmit
    timestamps are initialized to the originate timestamp value by the sender,
    setting all three timestamps to the same correct value. The destination IP
    address and originate timestamp are again encoded in the identifier and sequence
    number to detect modifications.
\end{enumerate}

Many implementation behaviors in~\S~\ref{sec:taxonomy} can be inferred from the
first, standard probe. For instance, the standard timestamp request
can determine a normal, lazy, non-UTC and little-endian implementation.
In order to classify a
device as stuck, both the standard and duplicate timestamp requests are
required. Two requests are needed in order to determine that the receive and
transmit timestamps remain fixed over time, and the inclusion of the
duplicate timestamp request ensures that the remote device is not simply
echoing the values in the receive and transmit timestamp fields of the
request.  Similarly, timestamp reflectors can be detected using the standard and
duplicate request responses.

The checksum-lazy behavior is detected via responses to the bad checksum request
type. The Linux \texttt{htons()} bug
behavior can be detected using the standard request and filtering for reply
timestamps with the two lower bytes set to zero. In order
to minimize the chance of false positives (\ie the correct time in ms from UTC
midnight is represented with the two lower bytes zeroed), we count
only destinations that match this behavior in responses from both the standard
and bad clock timestamp request types. 

To detect the unit precision of the timestamp reply fields, we leverage
the multiple requests sent to each target. Because we
know the time at which requests are transmitted, we
compare the time difference between the successive requests to a host and
classify them based on the inferred time difference from the replies.

Finally, we classify responsive devices by the reference by which they maintain
time.  We find many remote machines that observe nonstandard reference
times, but do not set the high order timestamp field bit.
A common alternative timekeeping methodology is to track the number of
ms elapsed since midnight local time. We detect local \tz
timekeepers by comparing the receive and transmit timestamps to the originate
timestamp in replies to the standard request. Receive and transmit timestamps
that differ from our correct originate timestamp by the number of ms  
for an existing \tz (within an allowable error discussed in \S~\ref{sec:correctness}) are determined to be keeping track of their local time.

Last, a small number of devices we encountered measured time relative to the
Unix epoch. Epoch-relative timestamps are detected in two steps: first, we
compare the epoch timestamp's date to the date in which we sent the request; if
they match, we determine whether the number of seconds elapsed since UTC
midnight in the reply is suitably close to the correct UTC time.

\subsection{Ground Truth}
\label{sec:gt}
To validate our inferences and understand the more
general behavior of popular operating systems and devices, we run 
\sundial against a variety of known systems; Table~\ref{tab:gt}
lists their \icmp timestamp reply behavior.

\begin{table}[t]
 \centering
 \caption{Ground Truth Classification of \icmp Timestamp Behaviors}
 \label{tab:gt}
 \vspace{-3mm}
 \scriptsize{
 \begin{tabular}{|l|c|c|}\hline
 \textbf{OS} & \textbf{Behavior} & \textbf{Notes} \\\hline\hline
 Windows 7 - 10 & Off by default & With Windows firewall off, lazy LE\\\hline
 Linux & Lazy & \\\hline
 Linux 3.18 (incl Android) & Lazy & htons() bug\\\hline
 Android kernel 3.10, 4.4+& Lazy & \\\hline 
 BSD & Lazy & \\\hline
 OSX & Unresponsive & \\\hline
 iOS & Off by default & \\\hline
 Cisco IOS/IOS-XE & Lazy & MSB set if NTP disabled, unset if enabled\\\hline
 JunOS & Lazy &  \\\hline
 \end{tabular}
 }
 \vspace{-6mm}
\end{table}

Apple desktop and mobile operating systems, macOS and iOS, 
both do not respond to \icmp timestamp messages by default. 
Initially, we
could not elicit any response from Microsoft Windows devices, until we disabled
Windows Firewall. Once disabled, the Windows device responds with
correct timestamps in little-endian byte order. This suggests that not only are
timestamp-responsive devices with little-endian timestamp replies Windows,
but it also worryingly indicates that its built-in firewall has been
turned off by the administrator. 

BSD and Linux devices respond with lazy timestamp replies, as
their source code indicates they should. 
JunOS and Android
respond like FreeBSD and Linux, on which they are based, respectively.
Of note, we built the Linux 3.18
kernel, which has the \texttt{htons()} bug described in \S~\ref{sec:buggy}; it
responded with the lower two bytes zeroed, as expected.  
This bug has made its way into Android, where we find devices running the 3.18
kernel exhibiting the same signature.

Cisco devices respond differently depending on whether they have enabled NTP.
NTP is not enabled by default on IOS; the administrator must manually enable the
protocol and configure the NTP servers to use. If NTP has not been enabled, we
observe devices setting the most significant bit, presumably to indicate that it
is unsure whether the timestamp is accurate, and filling in a UTC-based
timestamp with the remaining bits, according to its internal clock. 

\paragraph{Telnet Banner and CWMP GET Ground Truth}
To augment the ground truth we obtained from devices we were able to
procure locally, we leveraged \vfour Internet-wide Telnet banner- and CPE WAN
Management Protocol (CWMP) parameter-grabbing scans from 
\texttt{scans.io}~\cite{scansio}.  From October 3, 2018 scans, we search
banners (Telnet) and GET requests (CWMP) for IP addresses 
associated with known manufacturer strings.  We then probe 
these addresses with \sundial.

\begin{figure}[t]
\vspace{-4mm}
\centering
\begin{minipage}{.5\textwidth}
  \centering
 \resizebox{1.065\columnwidth}{!}{\includegraphics{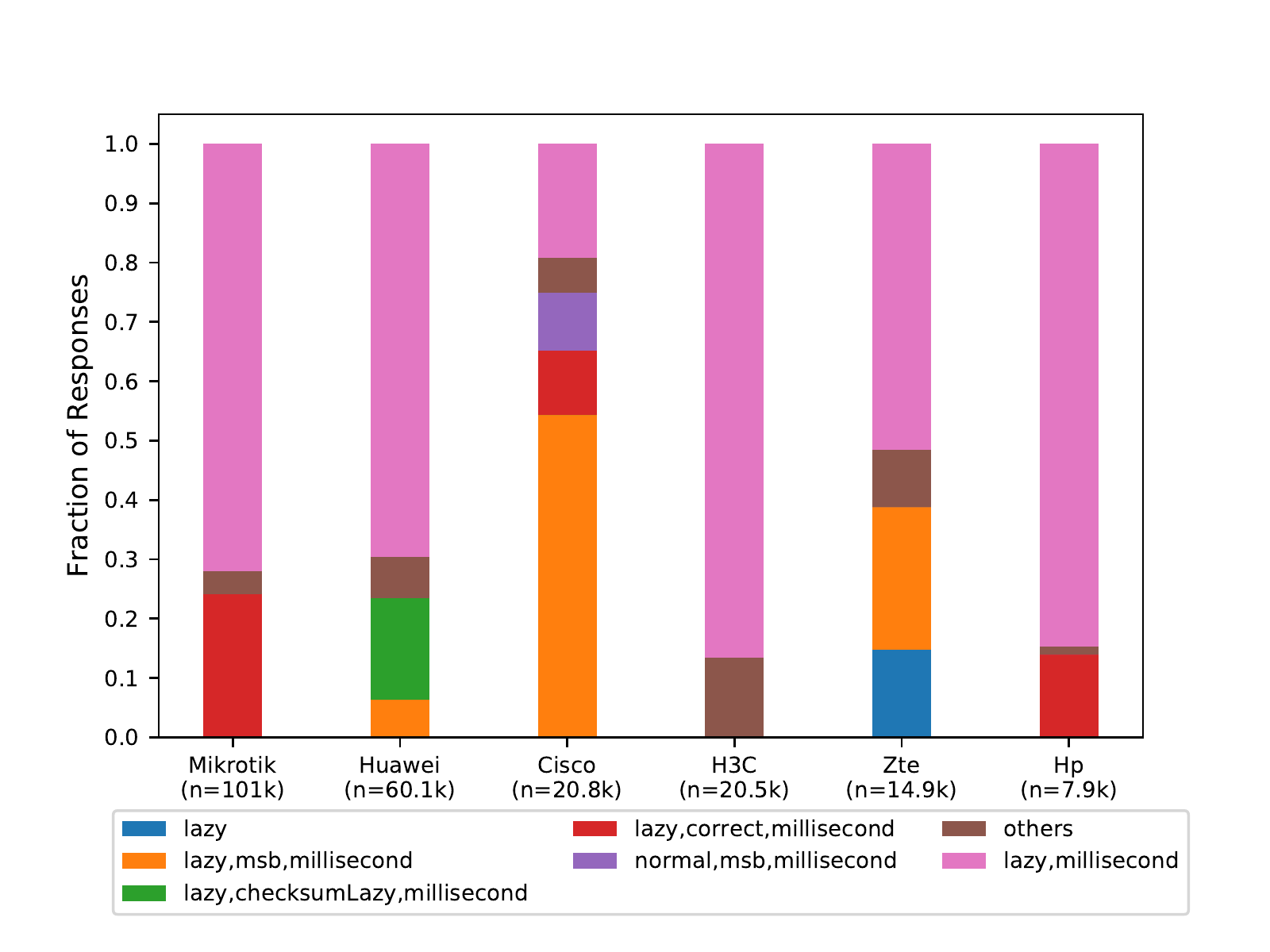}}
 \caption{Incidence of Fingerprints for Most Common Telnet Banner Manufacturers}
 \label{fig:telnetBar}
\end{minipage}%
\begin{minipage}{.5\textwidth}
  \centering
 \resizebox{1.065\columnwidth}{!}{\includegraphics{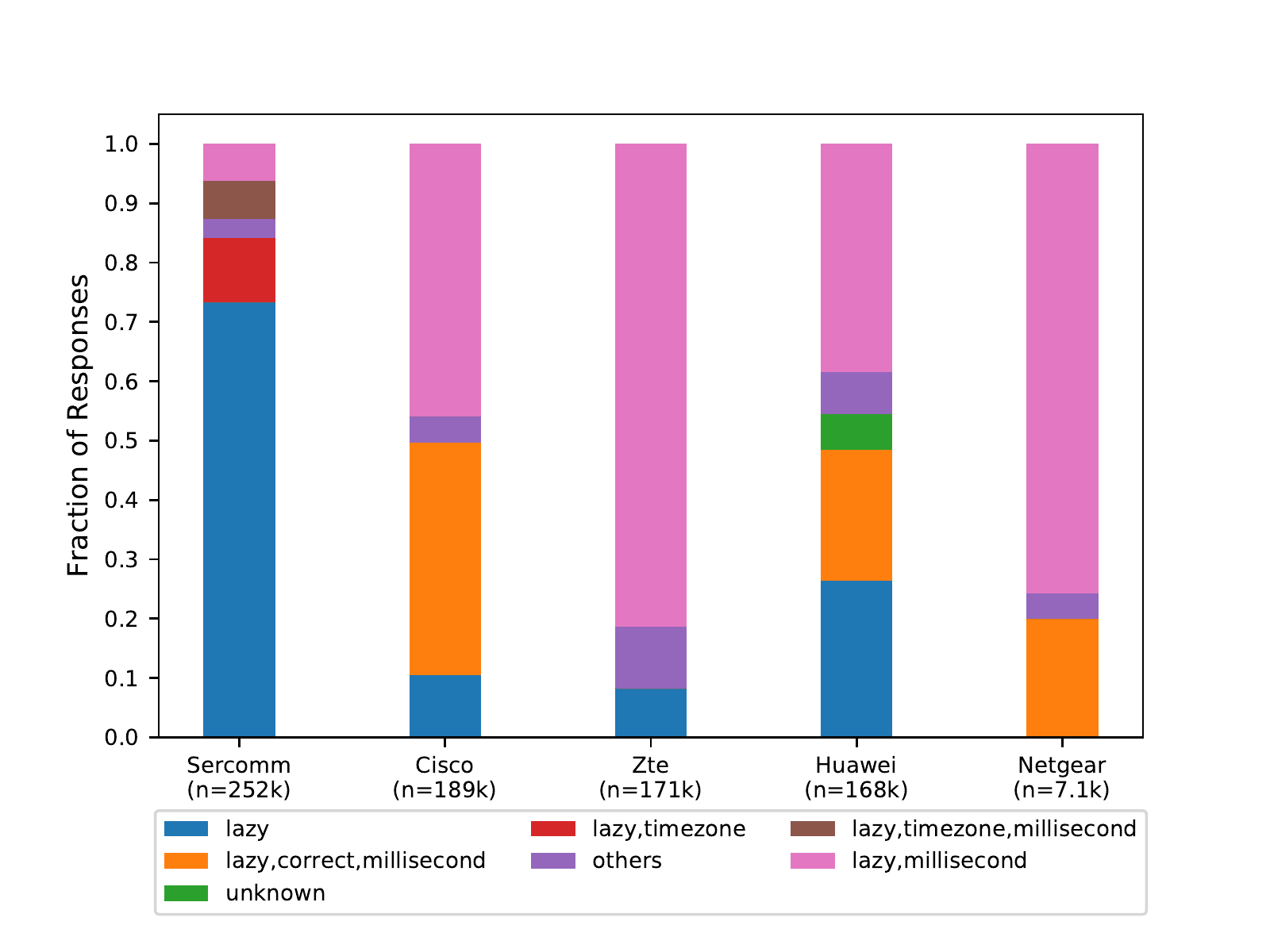}}
 \caption{Incidence of Fingerprints for Most Common CWMP Scan Manufacturers}
 \label{fig:cwmpBar}
\end{minipage}
\vspace{-6mm}
\end{figure}

Figure~\ref{fig:telnetBar} displays the most common fingerprints for a subset of
the manufacturers probed from \texttt{scans.io}'s Telnet banner-grab dataset,
while Figure~\ref{fig:cwmpBar} is the analogous CWMP plot. We note that
non-homogeneous behavior within a manufacturer's plot may be due to several
factors: different behaviors among devices of the same manufacturer, banner
spoofing, IP address changes, and middleboxes between the source and
destination. We provide further details regarding our use of the
\texttt{scans.io} datasets in Appendix B.

\subsection{Ethical Considerations}

Internet-wide probing invariably raises ethical concerns.  We
therefore follow the recommended guidelines for good Internet
citizenship provided in~\cite{durumeric2013zmap} to mitigate the
potential impact of our probing.  At a high-level, we
only send ICMP packets, which are generally considered less abusive
than \eg TCP or UDP probes that may reach active application services.
Further, our pseudo-random probing order is designed to distribute
probes among networks in time so that they do not appear as attack
traffic.  Finally, we make an informative web page accessible via the
IP address of our prober, along with instructions for opting-out.  In
this work, we did not receive any abuse reports or opt-out requests.

\section{Results}

On October 6, 2018, we sent four \icmp timestamp request messages as described
in \S~\ref{sec:requesttypes} from both of our vantage points to each of the 14.5
million target \vfour addresses in the ISI hitlist. We obtained at
least one \icmp timestamp reply message from 2,221,021 unique IP addresses in
$42,656$ distinct autonomous systems as mapped by Team Cymru's IP-to-ASN 
lookup service~\cite{cymru2008ip}. Our probing results are
publicly available~\cite{sundial}.

We classify the responses according to the
implementation taxonomy outlined in \S~\ref{sec:taxonomy} and
Table~\ref{tab:taxonomy}, the timekeeping behavior detailed in
\S~\ref{sec:timekeeping}, and the correctness of the timestamp reply according
to \S~\ref{sec:correctness}. Tables~\ref{tab:results} and
\ref{tab:timekeepingResults} summarize our results in tabular form; 
note that the implementation behavior categories are not mutually exclusive,
and the individual columns will sum to more than the total column, which is the
number of unique responding IP addresses. We received replies from approximately
11,000 IP addresses whose computed MD5 hashes as described in
\S~\ref{sec:requesttypes} indicated tampering of the source IP address,
originate timestamp, or id and sequence number fields; we discard these replies.

\subsection{Macro Behavior}

Lazy replies outnumber normal timestamp replies by a margin of
over 50 to 1. Because we had assumed the normal reply type would be the most
common, we investigated open-source operating systems' implementations of \icmp.
In both the Linux and BSD implementations, the receive timestamp is filled in
via a call to retrieve the current kernel time, after which this value
is simply copied into the transmit timestamp field.  Therefore, all BSD and
Linux systems, and their derivatives, exhibit the lazy timestamp reply
behavior. 

Normal hosts can appear lazy if the receive and transmit
timestamps are set within the same millisecond.  This ambiguity can
be resolved in part via multiple probes.  For instance,
Table~\ref{tab:results} shows that only $\sim$50\% of responders
classified as normal by one vantage are also marked normal by the
other.

The majority (61\%) of responding devices do not reply with timestamps within
200ms of our NTP-synchronized reference clock, our empirically-derived
correctness bound discussed in \S~\ref{sec:correctness}. Only 
$\sim$40\% of responding IP addresses fall into this category; notably, we detect
smaller numbers devices with correct clocks incorrectly implementing the
timestamp reply message standard. For example, across both vantage points we
detect thousands of devices whose timestamps are correct when interpreted as a
little-endian integer, rather than in network byte order. We discover one operating system that implements little-endian timestamps in
\S~\ref{sec:gt}. In another incorrect behavior that nevertheless indicates
a correct clock, some devices respond with the correct timestamp 
and the most significant bit set -- a behavior at odds with the
specification~\cite{rfc792} where the most significant bit indicates
a timestamp either not in ms, or 
the host cannot provide a timestamp referenced to UTC midnight.
In \S~\ref{sec:gt}, we discuss an
operating system that sets the most significant bit when its clock has not been
synchronized with NTP.

\begin{table}[t]
 \centering
  \caption{Timestamp Reply Implementation Behaviors (values do not sum to total)}
 \label{tab:results}
 \vspace{-3mm}
 \scriptsize{
 \begin{tabular}{|l|c|c|c||l|c|c|c|}\hline
   \textbf{Category} & \textbf{Boston} & \textbf{Both} &\textbf{San Diego} &
   \textbf{Category} & \textbf{Boston} & \textbf{Both} &\textbf{San Diego} 
\\\hline\hline
   Normal & 40,491& 19,819 &40,363 &
   Stuck&855 & 849 &873  \\\hline

   Lazy   &2,111,344 & 1,899,297 & 2,112,386 &
   Constant 0&547 & 546 &555 \\\hline

   Checksum-Lazy&28,074 &23,365&28,805 &
   Constant 1&200 & 199 & 207\\\hline

   Non-UTC&249,454 & 211,755  & 249,932 &
   Constant LE 1& 22& 19 &23 \\\hline

   Reflection&2,325 & 2,304 & 2,364 &
   \texttt{htons()} Bug &1,499 & 665 &1,536  \\\hline

   Correct  &850,787 & 803,314 &850,133&
   \Tz&33,317 & 23,464 & 33,762 \\\hline

   Correct LE & 11,127& 5,244 & 11,290 &
   Unknown & 38,495 & 11,865 & 32,956\\\hline

   Correct - MSB & 1,048&  386 & 973&
   & & & \\\hline

   Total & &&&&2,194,180 & 1,934,172 &2,189,524 \\\hline
 \end{tabular}
 }
 \vspace{-6mm}
\end{table}

Over 200,000 unique IPs ($>$10\% of each vantage point's total) respond with the
most significant bit set in the receive and transmit timestamps; those
timestamps that are otherwise correct are but a small population of those we
term Non-UTC due to the prescribed meaning of this bit in~\cite{rfc792}.
Some hosts and routers fall into this category due to the
nature of their timestamp reply implementation -- devices that mark the receive
and transmit timestamps with little-endian timestamps will be classified as
Non-UTC if the most significant bit of the lowest order byte is on, when the
timestamp is viewed in network byte order. Others, as described above, turn on
the Non-UTC bit if they have not synchronized with NTP. 

Another major category of non-standard implementation behavior of \icmp
timestamp replies are devices that report their timestamp relative to their
local \tz. Whether devices are programmatically reporting their local time
without human intervention, or whether administrator action is required to
change the system time (from UTC to local time) in order to effect this
classification is unclear. In either case, \tz timestamp replies allow us
to coarsely geolocate the responding device. We delve deeper into this
possibility in \S~\ref{sec:geolocation}.

\begin{table}[t]
 \centering
  \caption{Timestamp Reply Timekeeping Behaviors}
 \label{tab:timekeepingResults}
 \vspace{-3mm}
 \scriptsize{
   \begin{tabular}{|l|c|c|c|}\hline
   \textbf{Category} & \textbf{Boston} & \textbf{Both} &\textbf{San Diego}
\\\hline\hline
   Millisecond  &1,826,696 & 1,722,176& 1,866,529 \\\hline
   Second  & 47 & 37 &68 \\\hline
   Epoch  & 1 & 1 & 1 \\\hline
   Unknown Timekeeping & 367,436 &  211,958 & 322,926\\\hline
   Total &2,194,180 & 1,934,172 & 2,189,524 \\\hline
 \end{tabular}
 }
 \vspace{-6mm}
\end{table}

Finally, while most responding IP addresses are unsurprisingly classified as
using milliseconds as their unit of measure, approximately 14-16\% of IP addresses
are not (see Table~\ref{tab:timekeepingResults}). In order to determine what
units are being used in the timestamp, we subtract the time elapsed between the
standard timestamp request and duplicate timestamp request, both of which
contain correct originate timestamp fields. We then subtract the time elapsed
according to the receive and transmit timestamps in the timestamp reply
messages. If the difference of differences is less than 400ms (two times 200ms,
the error margin for one reply) we conclude that the remote IP is counting in
milliseconds. A similar calculation is done to find devices counting in seconds.
Several of the behavioral categories outlined in
\S~\ref{sec:taxonomy:structural} are included among the hosts with undefined
timekeeping behavior -- those whose clocks are stuck at a particular value and
those that reflect the request's receive and transmit timestamps into the
corresponding fields are two examples.  Others may be filling the reply timestamps with random values. 

\subsection{Timestamp Correctness}
\label{sec:correctness}

In order to make a final classification -- whether the remote host's clock is
correct or incorrect -- as well as to assist in making many of the
classifications within our implementation and timekeeping taxonomies that require a
correctness determination, we describe in this section our methodology for
determining whether or not a receive or transmit timestamp is correct.

\begin{figure}[t]
\vspace{-4mm}
\centering
\begin{minipage}{.5\textwidth}
  \centering
 \resizebox{1.0\columnwidth}{!}{\includegraphics{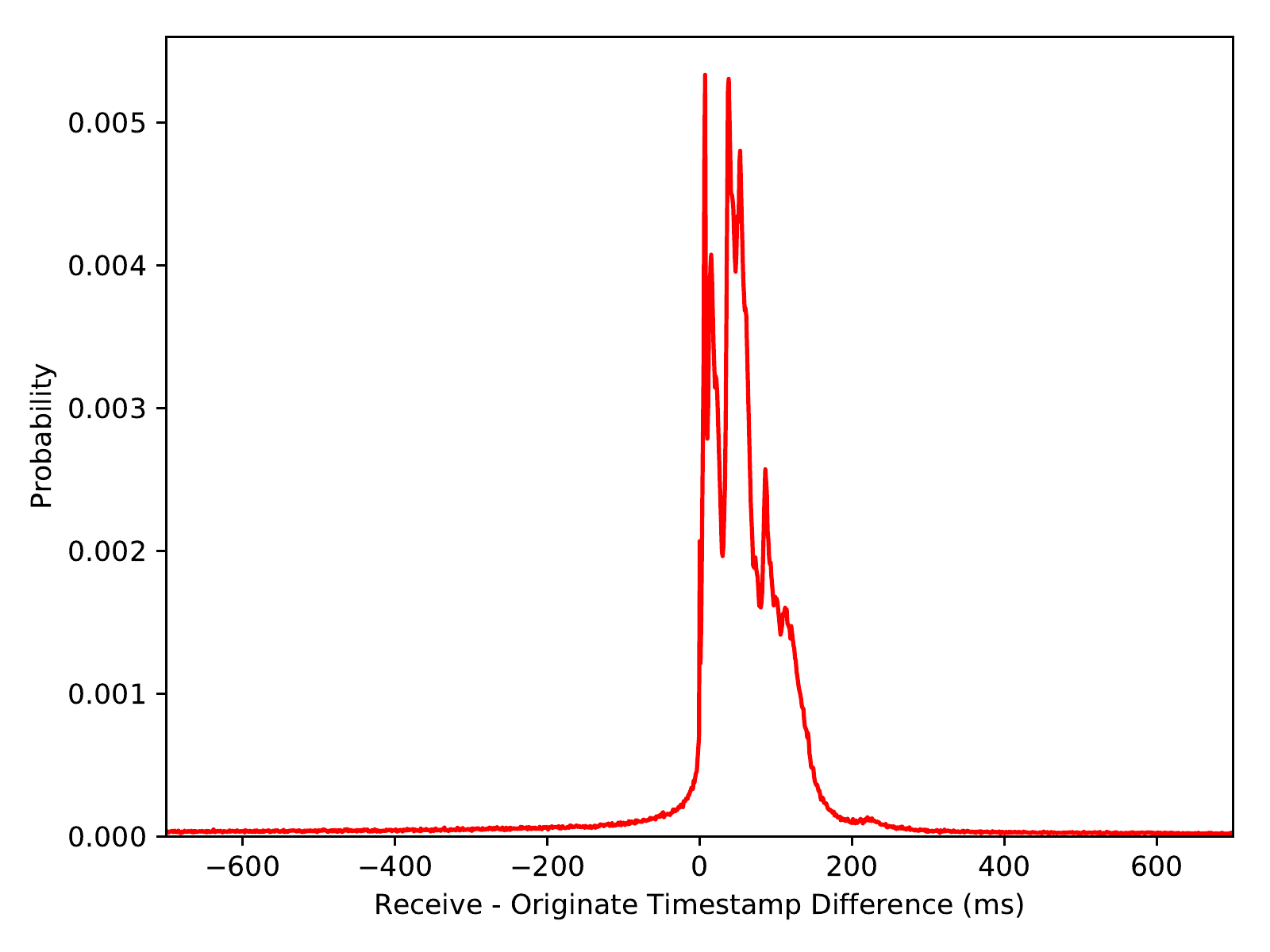}}
 \caption{Empirical \recv - \orig\xspace PMF}
 \label{fig:correctNarrow}
\end{minipage}%
\begin{minipage}{.5\textwidth}
  \centering
 \resizebox{1.0\columnwidth}{!}{\includegraphics{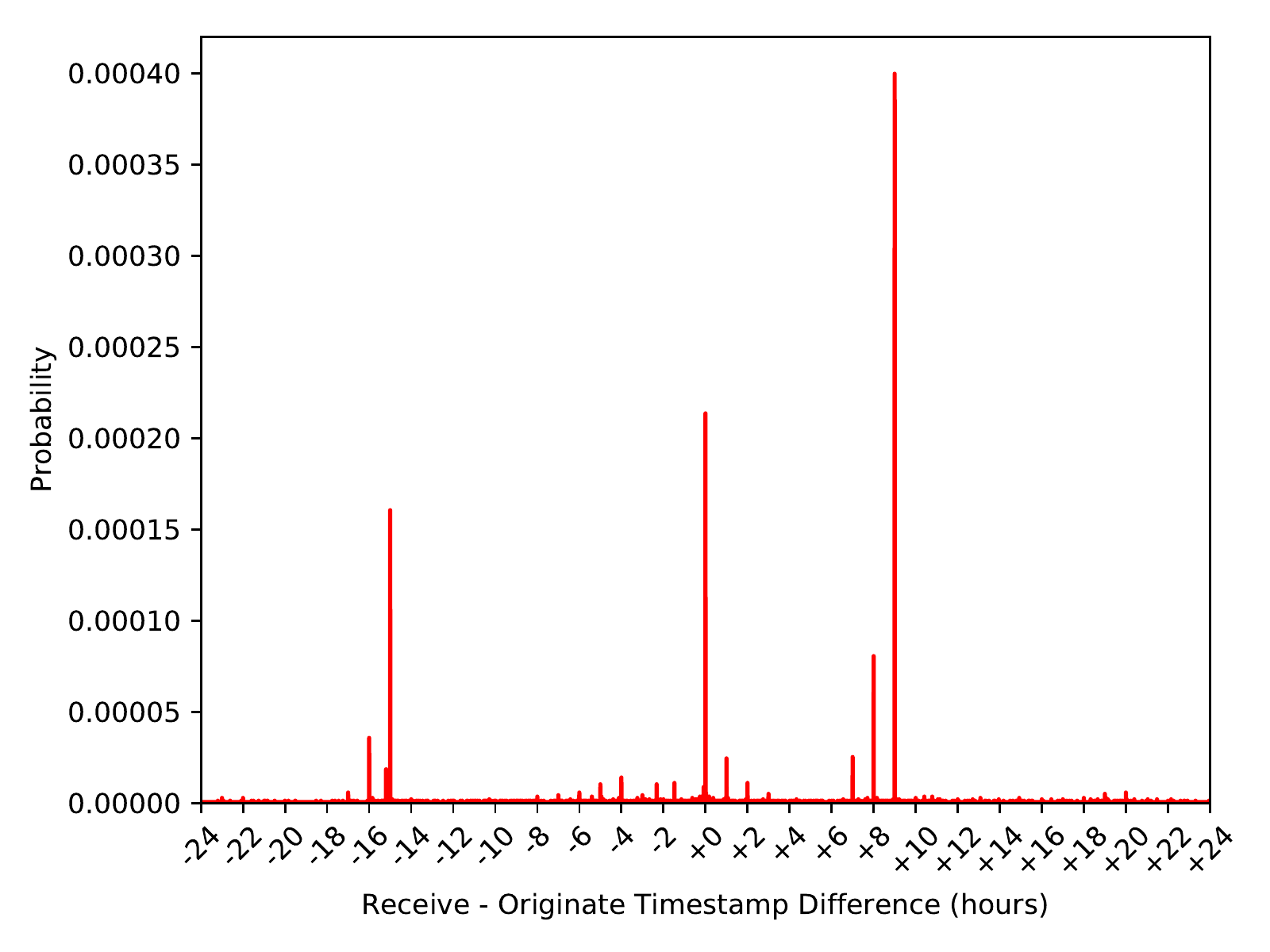}}
 \caption{Response error; note hourly peaks}
 \label{fig:timezone}
\end{minipage}
\vspace{-5mm}
\end{figure}

To account for clock drift and network delays, we aim to establish a margin of
error relative to a correctly marked originate timestamp, and consider receive
and transmit timestamps within that margin from the originate timestamp to be
correct. To that end, we plot the probability density of the
differences between the receive and originate timestamps from 2.2 million
timestamp replies generated by sending a single standard timestamp request to
each of 14.5 million IP addresses from the ISI
hitlist~\cite{fan2010selecting} in
Figure~\ref{fig:correctNarrow}.

Figure~\ref{fig:correctNarrow} clearly depicts a trough in the difference
probability values around 200ms, indicating that receive timestamps
greater than 200ms than the originate timestamp are less likely than
those between zero and 200ms. We reflect this margin about the
y-axis, despite the trough occurring somewhat closer to the origin on the
negative side.  Therefore, we declare a timestamp correct if it is within our
error margin of 200ms of the originate timestamp.

\subsection{Middlebox Influence}
To investigate the origin of some of the behaviors observed in
\S~\ref{sec:taxonomy} for which we have no ground truth implementations, we use
\texttt{tracebox}~\cite{detal2013revealing} to detect middleboxes. In
particular, we chose for investigation hosts implementing the reflection, lazy
with MSB set (but not counting milliseconds), and constant 0 behaviors, as we do
not observe any of these fingerprints in our ground truth dataset, yet there
exist nontrivial numbers of them in our Internet-wide dataset. 

In order to determine whether a middlebox may be responsible for these behaviors
for which we have no ground truth, we \texttt{tracebox} to a subset of 500
random IP addresses exhibiting them. For our purposes, we consider an IP address
to be behind a middlebox if the last hop modifies fields beyond the standard IP
TTL and checksum modifications, and DSCP and MPLS field alterations and
extensions.  Of 500 reflection IP addresses, only 44 showed evidence of being
behind a middlebox, suggesting that some operating systems implement the reflect
behavior and that this is a less common middlebox modification. The lazy with
MSB set (but non-ms counting) behavior, on the other hand, was inferred to be
behind a middlebox in 333 out of 500 random IP addresses, suggesting it is most
often middleboxes that are causing the lazy-MSB-set fingerprint. Finally, about
half of the constant 0 IP addresses show middlebox tampering in
\texttt{tracebox} runs, suggesting that this behavior is both an operating
system implementation of timestamp replies as well as a middlebox modification
scheme.

\subsection{Geolocation}
\label{sec:geolocation}

Figure~\ref{fig:timezone} displays the probability distribution of response
error, \eg $\recv - \orig$, after correct replies have been removed from the set
of standard request type responses.  While there is a level of uniform randomness,
we note the peaks at hour intervals.  We surmise that these represent hosts that
have correct time, but return a \emph{\tz-relative} response (in violation
of the standard~\cite{rfc792} where responses should be relative to UTC).  
The origin of \tz-relative responses may be a non-conformant
implementation. Alternatively, these responses may simply be an
artifact of non-NTP synchronized machines where the administrator
instead sets the localtime correctly, but incorrectly sets the
\tz.  In this case, the machine's notion of UTC is incorrect,
but incorrect relative to the set \tz.
Nevertheless, these \tz-relative responses effectively leak
the host's \tz. We note the large spike in the +9 \tz, which
covers Japan and South Korea; despite the use of \texttt{nmap}'s OS-detection
feature, and examining web pages and TLS certificates where available, we could
not definitively identify a specific device manufacturer or policy underpinning
this effect.

To evaluate our ability to coarsely geolocate IP addresses reporting a
\tz-relative timestamp, we begin with $\sim$34,000 IP addresses in this
category obtained by sending a single probe to every hitlist IP from our Boston
vantage.  Using the reply timestamps, we compute the remote host's local
\tz offset relative to UTC to infer the host's \tz.  We then compare
our inferred \tz with the \tz reported by the MaxMind GeoLite-2
database~\cite{maxmind}. 

For each IP address, we compare the MaxMind \tz's standard time UTC-offset
and, if applicable, daylight saving time UTC offset, to the timestamp-inferred
offset. Of the 34,357 IP addresses tested, 32,085 (93\%) correctly matched
either the standard \tz UTC offset or daylight saving UTC offset, if the
MaxMind-derived \tz observes daylight saving time. More specifically,
18,343 IP addresses had timestamp-inferred \tz offsets that matched their
MaxMind-derived \tz, which did not observe daylight saving time. 11,188 IP
addresses resolved to a MaxMind \tz, whose daylight saving time offset
matched the offset inferred from the timestamp. 2,554 IP addresses had
timestamp-inferred UTC offsets that matched their MaxMind-derived standard
time offset for \tzp that do observe daylight saving time. Of the inferred
UTC-offsets that were not correct, 1,641 did not match either the standard time
offset derived from MaxMind, or the daylight saving time offset, if it existed,
and 631 IP addresses did not resolve to a \tz in MaxMind's free database.

\section{Conclusions and Future Work}
We observe a wide variety of implementation behavior of the \icmp timestamp
reply type, caused by timestamps' lack of a modern use but continued requirement
to be supported. In particular, we are able to uniquely fingerprint the behavior
of several major operating systems and kernel versions, and geolocate Internet
hosts to \tz accuracy with $>$90\% success.

As future work, we intend to exhaustively scan and classify the \vfour Internet,
scan a subset with increased frequency over a sustained time
period, and to do so many vantage points.  We further plan to 
integrate the OS-detection capabilities we uncover in this work into
\texttt{nmap}, and 
add \texttt{tracebox}
functionality to \sundial in order to better detect middlebox tampering with
\icmp timestamp messages.  

\section*{Acknowledgments}
\vspace{-1mm}
We thank Garrett Wollman, Ram Durairajan, and Dan Andersen for 
measurement infrastructure, our shepherd Rama Padmanabhan, and the anonymous reviewers for 
insightful feedback.
Views and conclusions are those of the authors and
not necessarily
those of the U.S.\ government.

\newpage
\small{
\bibliographystyle{splncs04}
\bibliography{sundial}
}

\newpage
\section*{Appendix A: Linux \texttt{htons()} Bug}
\label{sec:buggy}

While investigating the source code of open-source operating systems'
implementation of \icmp timestamps, we observed a flaw that
allows fine-grained fingerprinting of the Linux kernel version 3.18. The
specific bug that allows this fingerprinting was introduced in March 2016. An
update to the Internet timestamp generating method in \texttt{af\_inet.c}
errantly truncated the 32-bit timestamp to a 16-bit short via a call to the C
library function \texttt{htons()} rather than \texttt{htonl()}. When this
incorrect 16-bit value is placed into the 32-bit receive and transmit timestamp
fields of a timestamp reply, it causes the lower two bytes to be zero and
disables the responding machine's ability to generate a correct reply timestamp
at any time other than midnight UTC. This presents a unique signature of devices
running the Linux kernel built during this time period. In order to identify
these devices on the Internet, we filter for \icmp timestamp replies containing
receive and transmit timestamp values with zeros in the lower two bytes when
viewed as a 32-bit big-endian integer. While devices that are correctly
implementing \icmp timestamp replies will naturally reply with timestamps
containing zeros in the lower two bytes every 65,536 milliseconds, the
probability of multiple responses containing this signature drops rapidly as the
number of probes sent increases.

Being derived directly from the Linux kernel, the 3.18 version of the Android
kernel also includes the flawed \texttt{af\_inet.c} implementation containing
the same \texttt{htons()} truncation, allowing for \icmp timestamp
fingerprinting of mobile devices as well. 

While Linux 3.18 reached its end of life~\cite{linuxkernelarchives} in 2017, we
observe hosts on the Internet whose signatures suggest this is the precise
version of software they are currently running. Unfortunately, this presents an
adversary with the opportunity to perform targeted attacks.

\section*{Appendix B: \texttt{scans.io} Ground Truth}
\label{sec:scans}

We use Telnet and CWMP banners in public \texttt{scans.io} as a 
source of ground truth.
It is possible to override
the default text of these protocol banners, 
and recognize that this is
a potential source of error. However, we examine the manufacturer counts in
aggregate under the assumption that most manufacturer strings are legitimate. We
believe it unlikely that users have modified their CWMP configuration on their
customer premises equipment to return an incorrect manufacturer.

Parsing the Telnet and CWMP scans for strings containing the names of major
network device manufacturers provided over two million unique IP addresses.
Table~\ref{tab:scancount} summarizes the results; note that for some
manufacturers (\eg Arris) approximately the same number of IPs were discovered
through the Telnet scan as the CWMP scan, for others (\eg Cisco and Huawei) CWMP
provided an order of magnitude greater number of IPs, and still others (\eg
Mikrotik and Netgear) appeared in only one of the two protocol scans. Note that
these numbers are not the number of timestamp-responsive IP addresses denoted by
$n$ in Figures~\ref{fig:telnetBar} and \ref{fig:cwmpBar}.

 \begin{table}[t]
   \centering
   \caption{Unique IP Addresses per Manufacturer for Each Scan}
  \label{tab:scancount}
   \vspace{-2mm}
   \scriptsize{
  \begin{tabular}{|l|c|c|}\hline
     Manufacturer & Telnet Count & CWMP Count \\\hline\hline
     Arris &8,638 & 5,281  \\\hline
     Cisco &29,135 & 1,298,761   \\\hline
     H3C &80,445 & -  \\\hline
     HP &24,027 & -  \\\hline
     Huawei &170,710 & 2,377,079  \\\hline
     Mikrotik & 190,484 & -  \\\hline
     Netgear & - & 17,723   \\\hline
     Sercomm & - & 899,492   \\\hline
     Ubiquiti &598 & -  \\\hline
     Zhone &6,999 & -  \\\hline
     ZTE &17,972 & 560,177  \\\hline
     Zyxel &5,902 & -  \\\hline
  \end{tabular}
   }
  \vspace{-3mm}
 \end{table}

With the IP addresses we obtained for each manufacturer, we then run
\texttt{sundial} to each set in order to elicit timestamp reply fingerprints and
determine whether different manufacturers tend to exhibit unique reply
behaviors. Figures~\ref{fig:telnetBar} and \ref{fig:cwmpBar} display the
incidence of timestamp reply fingerprints for a subset of the manufacturers we
probed, and provide some interesting results that we examine here in greater
detail. 

No manufacturer exhibits only a singular behavior. We
attribute this variety within manufacturers to changes in their implementation
of timestamp replies over time, different implementations among different
development or product groups working with different code bases, and the
incorporation of outside implementations inherited through 
acquisitions and mergers.

Second, we are able to distinguish broad outlines of different manufacturers
based on the incidence of reply fingerprints. In Figure~\ref{fig:telnetBar}, we
note that among the top six manufacturers, only Huawei had a significant number
of associated IP addresses ($\sim$10\%) that responded with the checksum-lazy
behavior. More than half of the Cisco IP addresses from the Telnet scan
exhibited the lazy behavior with the most significant bit set while counting
milliseconds, a far greater proportion than any other manufacturer. Also
noteworthy is that none of the manufacturers represented in the Telnet scan
exhibits large numbers of correct replies. In our Telnet data, Mikrotik devices
responded with a correct timestamp reply roughly 25\% of the time, a higher
incidence than any other manufacturer.  This suggests that perhaps certain
Mikrotik products have NTP enabled by default, allowing these devices to obtain
correct time more readily than those that require administrator interaction. Our
CWMP results in Figure~\ref{fig:cwmpBar} demonstrate the ability to distinguish
manufacturer behavior in certain cases as well, we note the $>70$\% of Sercomm
devices that exhibit only the lazy behavior, as well as Sercomm exhibiting the
only \tz-relative timekeeping behavior among the CWMP manufacturers.

Finally, we note differences between the protocol scans among IP addresses that
belong to the same manufacturer. Cisco, Huawei, and ZTE appear in both protocol
results in appreciable numbers, and are represented in both figures in
\S~\ref{sec:gt}. Although Cisco devices obtained from the Telnet scan
infrequently ($\sim$10\%) respond with correct timestamps, in the CWMP data
the proportion is nearly 40\%. Huawei devices from the Telnet data are generally
lazy responders that count in milliseconds, however, this same behavior occurs
only half as frequently in the CWMP data. Further, the fingerprint consisting
solely of the lazy behavior represents nearly a quarter of the CWMP Huawei
devices, while it is insignificant in the Telnet Huawei data. While the
differences between the Telnet and CWMP data are less pronounced for ZTE, they
exist as well in the lack of appreciable numbers of ZTE devices setting the most
significant bit in replies within the CWMP corpus.

\section*{Appendix C: \Tz-Relative Behavior}
\begin{table}[t]
 \centering
 \label{tab:timezones}
  \caption{Inferred UTC-Offsets from Timestamp Replies}
 \vspace{-2mm}
 \scriptsize{
 \begin{tabular}{|l|c|c|c|c|c|c|c|c|c|c|c|c|c|c|c|}\hline
   \textbf{UTC Offset} &\textbf{-12}&\textbf{-11}&\textbf{-10}&\textbf{-9}&\textbf{-8}&\textbf{-7}&\textbf{-6}&\textbf{-5}&\textbf{-4}&\textbf{-3.5}&\textbf{-3}&\textbf{-2}&\textbf{-1}&\textbf{1}&\textbf{2}\\\hline
   Count & 73&1&7&3&386&476&666&1,763&2,660&2&246&228&5&7,215&1,819\\\hline
   \textbf{UTC Offset}
   &\textbf{3}&\textbf{3.5}&\textbf{4}&\textbf{4.5}&\textbf{5}&\textbf{5.5}&\textbf{6}&\textbf{6.5}&\textbf{7}&\textbf{8}&\textbf{9}&\textbf{9.5}&\textbf{10}&\textbf{11}&\\\hline
   Count &449&8&62&3&87&17&14&13&565&3,496&13,861&6&215&11&\\\hline
 \end{tabular}
 }
\end{table}

Figure~\ref{fig:timezone} displays the probability mass function of the
differences between the receive and originate timestamps for a \texttt{sundial}
scan conducted on 9 September 2018 from the Boston vantage after responses with
correct timestamps have been removed. Discernible peaks occur at many of the
hourly intervals representing \tz-relative responders, rising above a base level
of randomness. The hourly offsets in Figure~\ref{fig:timezone} may need to be
normalized to the range of UTC timezone offsets, however. For example, depending
on the originate timestamp value, a responding host's receive timestamp at a UTC
offset of $+9$ may appear either nine hours ahead of the originate timestamp, or
$15$ hours behind, as $-15\ \equiv 9 (\textrm{mod}\ 24)$. In
Figure~\ref{fig:timezone} we see large spikes at both $+9$ and $-15$ hours, but
in reality these spikes represent the same \tz. 

We identify \tz-relative responses systematically by computing the local time in
milliseconds for each of the UTC-offsets detailed in Table~\ref{tab:timezones},
given the originate timestamp contained in the timestamp response. We then
compare each candidate local \tz's originate timestamp to the receive timestamp
in the reply. If the candidate originate timestamp is within the 200ms
correctness bound established in \S~\ref{sec:correctness}, we classify the IP
address as belonging to the \tz that produced the correct originate timestamp.
Table~\ref{tab:timezones} details the number of \tz-relative responders we
identified during the 9 September \texttt{sundial} scan. 

\end{document}